\documentstyle[aps,epsfig,prl]{revtex} 

\newcommand{\klong}{{{K_L}}}

\newcommand{\kleeg}  {{\klong \to e^+ e^- \gamma }}
\newcommand{\kleegg} {{\klong \to e^+ e^- \gamma \gamma}}

\newcommand{\kleegpg} {{\klong \to e^+ e^- \gamma (\gamma)}}
\newcommand{\kleeggc}{{\klong \to e^+ e^- \gamma \gamma, 
E^*_\gamma > 5\,\mbox{MeV}}}
\newcommand{\pid}      {{\pi^{0} \to e^+ e^- \gamma}}
\newcommand{\ket}      {{\klong \to \pi^{\pm} e^{\mp} \nu}}
\newcommand{\klpiee}   {{\klong \to \pi^0 e^+ e^-}}
\newcommand{\klpipi}  {{K_L \to \pi^0 \pi^0}}
\newcommand{\kpipid}  {{K \to \pi^0 \pi^0_{D}}}

\newcommand{\kpipipid}  {{K_L \to \pi^0 \pi^0 \pi^0_{D}}}
\newcommand{\akstar}{{\alpha_{K^*}}}
\newcommand{\aeff}{{\alpha_{\rm Eff}}}

\newcommand{\egkcm}{{E_\gamma^*}}
\newcommand{\mintheg}{{\theta_{min}}}

\newcommand{\meegg}{{m_{ee\gamma \gamma}}}

\newcommand{\mevoc}{\mbox{MeV/c}}
\newcommand{\gevoc}{\mbox{GeV/c}}
\newcommand{\mevocc}{{\mbox{MeV/c}^2}}

\begin{document}

%
\title{A Measurement of the Branching Ratio of $\kleegg$}
\date{\today}


\author{
The KTeV Collaboration \\
A.~Alavi-Harati$^{12}$,
T.~Alexopoulos$^{12}$,
M.~Arenton$^{11}$,
K.~Arisaka$^2$,
S.~Averitte$^{10}$,
A.R.~Barker$^5$,
L.~Bellantoni$^7$,
A.~Bellavance$^9$,
J.~Belz$^{10}$,
R.~Ben-David$^{7}$,
D.R.~Bergman$^{10}$,
E.~Blucher$^4$, 
G.J.~Bock$^7$,
C.~Bown$^4$, 
S.~Bright$^4$,
E.~Cheu$^1$,
S.~Childress$^7$,
R.~Coleman$^7$,
M.D.~Corcoran$^9$,
G.~Corti$^{11}$, 
B.~Cox$^{11}$,
M.B.~Crisler$^7$,
A.R.~Erwin$^{12}$,
R.~Ford$^7$,
P.M.~Fordyce$^5$,
A.~Glazov$^4$,
A.~Golossanov$^{11}$,
G.~Graham$^4$, 
J.~Graham$^4$,
K.~Hagan$^{11}$,
E.~Halkiadakis$^{10}$,
K.~Hanagaki$^8$,  
S.~Hidaka$^8$,
Y.B.~Hsiung$^7$,
V.~Jejer$^{11}$,
D.A.~Jensen$^7$,
R.~Kessler$^4$,
H.G.E.~Kobrak$^{3}$,
J.~LaDue$^{5,\dagger}$,
A.~Lath$^{10}$,
A.~Ledovskoy$^{11}$,
P.L.~McBride$^7$,
P.~Mikelsons$^{5}$,
E.~Monnier$^{4,*}$,
T.~Nakaya$^7$,
K.S.~Nelson$^{11}$,
H.~Nguyen$^7$,
V.~O'Dell$^7$, 
M.~Pang$^7$, 
R.~Pordes$^7$,
V.~Prasad$^4$, 
C.~Qiao$^4$,
B.~Quinn$^4$,
E.J.~Ramberg$^7$, 
R.E.~Ray$^7$,
A.~Roodman$^4$, 
M.~Sadamoto$^8$, 
S.~Schnetzer$^{10}$,
K.~Senyo$^8$, 
P.~Shanahan$^7$,
P.S.~Shawhan$^4$,
W.~Slater$^2$,
N.~Solomey$^4$,
S.V.~Somalwar$^{10}$, 
R.L.~Stone$^{10}$, 
I.~Suzuki$^8$,
E.C.~Swallow$^{4,6}$,
S.A.~Taegar$^1$,
R.J.~Tesarek$^{10}$, 
G.B.~Thomson$^{10}$,
P.A.~Toale$^5$,
A.~Tripathi$^2$,
R.~Tschirhart$^7$, 
Y.W.~Wah$^4$,
J.~Wang$^1$,
H.B.~White$^7$, 
J.~Whitmore$^7$,
B.~Winstein$^4$, 
R.~Winston$^4$, 
T.~Yamanaka$^8$,
E.D.~Zimmerman$^4$
\vspace*{0.1in}
}

\address{
$^1$ University of Arizona, Tucson, Arizona 85721 \\
$^2$ University of California at Los Angeles, Los Angeles, California 90095 \\
$^{3}$ University of California at San Diego, La Jolla, California 92093 \\
$^4$ The Enrico Fermi Institute, The University of Chicago, 
Chicago, Illinois 60637 \\
$^5$ University of Colorado, Boulder, Colorado 80309 \\
$^6$ Elmhurst College, Elmhurst, Illinois 60126 \\
$^7$ Fermi National Accelerator Laboratory, Batavia, Illinois 60510 \\
$^8$ Osaka University, Toyonaka, Osaka 560 Japan \\
$^9$ Rice University, Houston, Texas 77005 \\
$^{10}$ Rutgers University, Piscataway, New Jersey 08855 \\
$^{11}$ The Department of Physics and Institute of Nuclear and 
Particle Physics, University of Virginia, 
Charlottesville, Virginia 22901 \\
$^{12}$ University of Wisconsin, Madison, Wisconsin 53706 \\
$^{*}$ On leave from C.P.P. Marseille/C.N.R.S., France \\
$^{\dagger}$ To whom correspondence should be addressed. \\
jladue@fnal.gov.
}
\maketitle

\begin{abstract}
We report on a study of the decay $\kleegg$ carried out as a part
of the KTeV/E799 experiment at Fermilab.
The 1997 data yielded a sample of 1543 events,
including an expected background of $56 \pm 8$ events.
An effective form factor
was determined from the observed distribution
of the $e^+e^-$ invariant mass. Using this form factor in the calculation
of the detector acceptance, the branching ratio was measured to be
${\mathcal B}(\kleeggc) =
(5.84 \pm 0.15 {\rm ~(stat)} \pm 0.32 {\rm ~(sys)})\times 10^{-7} $.
\end{abstract}

%
%
%
%

\twocolumn

The process $\kleegg$ occurs mainly through kaon Dalitz decay 
($\kleeg$) with an internally radiated photon.  
As such, measurements of the radiative decay can test 
radiative correction calculations and probe the form factor of the 
$\klong \gamma \gamma^*$ vertex.
The study of $\kleegg$ is also important because it is a 
significant background to searches for $\klpiee$,
a mode which has an important direct $CP$-violating component and
was a primary focus of the KTeV/E799 rare-decay program \cite{Leo}.

The best existing measurements of the branching ratio ${\mathcal B}(\kleeggc)$
were made by the E799-I and NA31 experiments, which
measured 
$(6.5 \pm 1.2 {\rm ~(stat)} \pm 0.6 {\rm ~(sys)})\times 10^{-7}$ \cite{e799i}
and
$(8.0 \pm 1.5 {\rm ~(stat)} ^{+1.4}_{-1.2} {\rm ~(sys)})\times 10^{-7}$
\cite{NA31}, respectively,
based on signals of 58 and 40 events. 
This Letter presents a branching ratio measurement with much higher
precision, 
using data collected by the KTeV/E799 experiment at Fermilab in 1997.

\begin{figure}
\epsfig{file=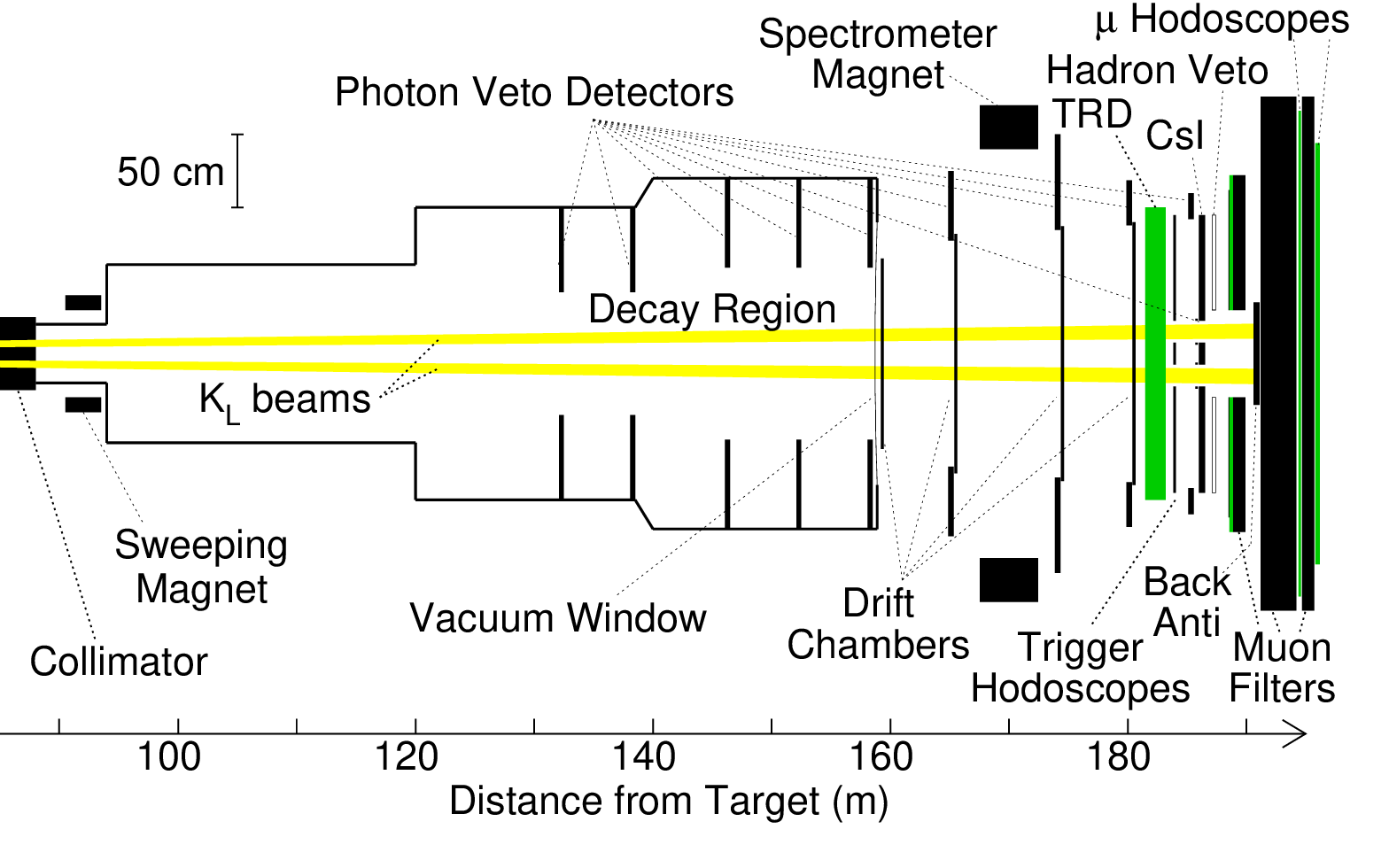,width=3.5in}
\caption{The KTeV/E799 detector configuration for rare decay studies.}
\label{fig:detector}
\end{figure}

Figure \ref{fig:detector} shows a plan view of the KTeV/E799 detector.
The Tevatron at Fermilab made an 800\,GeV proton beam, 
typically delivering $3.5 \times 10^{12}$ protons per minute.
The beam was incident on a BeO target.
Photons from the target were converted by a lead absorber immediately 
downstream.
Charged particles were then removed with magnetic sweeping.
Collimators defined two neutral beams
that entered the KTeV/E799 apparatus 94\,m downstream of the target. 
A 65\,m vacuum decay region extended to the first drift chamber.  
The charged-particle spectrometer consisted of a dipole magnet 
and four drift chambers 
(two upstream and two downstream of the magnet)
with $\sim100\mu \rm m$ position resolution 
in both horizontal and vertical directions. 
The magnetic field imparted a $205\,\mevoc$ horizontal momentum kick. 
The spectrometer had a momentum resolution of
$\sigma(P)/P \sim 0.38\% \oplus 0.016\% P$, where $P$\ is in \gevoc.

Photons were detected using an electromagnetic calorimeter, 
which consisted of 3100 pure CsI crystals, each 50\,cm long \cite{CsI}. 
Crystals in the central section of the calorimeter 
had a cross-sectional area of \mbox{$2.5 \times 2.5\,\mbox{cm}^2$}, 
and those in the outer region had a \mbox{$5 \times 5\,\mbox{cm}^2$} area.
The calorimeter's energy resolution for photons was
$\sigma(E)/E \sim 0.45\% \oplus 2\%/\sqrt{E}$, where $E$\ is in GeV.
The position resolution was about 1\,mm.
By comparing spectrometer momentum to calorimeter energy,
$e^\pm$ could be distinguished from $\pi^\pm$.
Nine photon veto assemblies (lead scintillator sandwiches)
detected particles leaving the fiducial volume. 

Additional $e^\pm/\pi^\pm$ separation was provided by eight transition
radiation detector (TRD) stations located downstream of the
last drift chambers \cite{GGHM}.
Each TRD station consisted of a polypropylene felt radiator 
followed by two planes of multiwire proportional chamber (MWPC)
filled with an 80/20\% mixture of Xe/CO$_{2}$ gas.
The pulse heights from the 16 MWPC planes,
compared with the pion pulse height distributions from a sample of
$\ket$ decays,
were used to calculate the confidence level 
that a given track was a pion.
Pion rejection factors over 200:1 were possible 
with a 90\% electron efficiency.

The trigger system required hits in two scintillation hodoscopes just upstream
of the CsI, as well as drift chamber hits consistent with two coincident 
charged particles
passing through the detector.  
The trigger used in this study 
counted the number of isolated clusters of in-time energy 
over 1\,GeV with a special processor \cite{HCC}; 
at least four such clusters were required.
The total energy deposited in the calorimeter had to be greater than 28\,GeV.  
The scintillator hodoscope behind a
lead wall downstream of the calorimeter 
vetoed events with hadronic showers in the calorimeter, 
and events with activity in the photon veto system were similarly discarded.
Events which passed the hardware trigger requirements 
were read out into on-line CPUs, 
which performed basic event reconstruction 
and applied a few loose event-topology and particle-identification cuts 
to select events to be written to tape.

During offline event reconstruction, 
each event was required to contain two tracks from
oppositely-charged particles originating from a common vertex.  
The reconstructed vertex had to be between 96\,m and 158\,m from the target.
The tracks had to be at least 1\,cm apart at the most upstream drift chamber, 
with an opening angle between the tracks of at least 2.24\,mrad.  
To select electrons, each track was required 
to point to a calorimeter cluster with an energy equal, 
within $\pm5\%$, 
to the track momentum measured by the spectrometer.  
The probability that a track was caused by a pion, 
formed by combining TRD chamber signals,
was required to be less than $4\%$ for each track.
The total momentum of all decay products had to be
under 216$\gevoc$ and the total energy greater than 33 $\rm GeV$.
Events with a total of four clusters were analyzed as $\kleegg$ candidates.
Events with a total of five clusters were
analyzed as $\kpipid$ decays, 
where $\pi^0_D$ denotes a pion decaying to $e^+e^-\gamma$.  
The latter mode was used both to check the simulation 
and to measure the total number of $K_L$ decays in the data sample.

In addition to the $\kleegg$ signal mode, 
several potential background processes were considered.  
The $\kleeg$ mode was a background when an additional photon was present.  
This photon could have been radiated by an $e^\pm$
while passing through matter in the detector (``external'' radiation), 
or it could have been reconstructed from coincidental activity 
in the calorimeter not associated with the decay (``accidental'' activity).  
$\kpipid$ was a background when one photon was missed.  
Similarly, $\kpipipid$ could have been a background 
but was suppressed 
because missing three photons generally resulted in a reconstructed mass 
well below the $K_L$ mass.  Background from $\ket$ plus two additional 
accidental
photons with the $\pi^\pm$ misidentified as $e^\pm$ by the calorimeter, 
was eliminated using the TRDs.

Monte Carlo (MC) simulations of the detector were performed
to calculate the acceptances for $\kleegg$ and $\kpipid$
and the misidentification rates for $\kleeg$, $\kpipid$ and $\ket$.
The effect of accidental clusters was simulated
by overlaying on the MC events with data events
taken with a random trigger that had a rate 
proportional to  the beam intensity.
The finite angle of external radiation was simulated.
Simulations of Dalitz decays included
radiative corrections to ${\mathcal O}(\alpha_{EM}^2)$
based on the work of Mikaelian and Smith \cite{MS}.
The radiative correction simulation also quantitatively predicted the ratios
$\Gamma(\kleeg)/\Gamma(K_L\to\gamma\gamma)$ and
$\Gamma(\kleegg)/\Gamma(K_L\to\gamma\gamma)$ 
and kinematic distributions 
such as photon energy in the $\klong$ center-of-mass, $\egkcm$.
The simulations of $\kleegg$ and $\kleeg$ used
the Bergstr\"{o}m, Mass\'{o}, and Singer (BMS) form factor \cite{BMS},
which includes a parameter $\akstar$ describing the relative strengths of 
the intermediate vector and pseudoscalar meson amplitudes
in the $K_L \rightarrow \gamma\gamma^*$ vertex.

The number of $\klong$ decays was measured by the $\kpipid$ decay mode,
as in the $\klpiee$ analysis~\cite{Leo}.  There were
$(2.642\pm0.015 {\rm ~(stat)} \pm 0.024 {\rm ~(sys)} \pm 0.091 {\rm ~(BR)})
\times10^{11}$ decays in the fiducial region, where the third term is from
uncertainties on the branching ratios of $\klpipi$ and $\pid$~\cite{PDGN}.
%

Several cuts on reconstructed quantities were made
on events with exactly four calorimeter clusters
to identify $\kleegg$ signal candidates.
First, to ensure that candidate events did not include
radiated photons below the nominal infrared cutoff of 5 MeV, 
$\egkcm$ was calculated for each photon 
and was required to be at least 8 MeV 
(to allow for finite detector resolution).  
Since the MC sample 
was subjected to the same requirement,
the acceptance correction procedure yielded a branching ratio
valid for $\egkcm > 5\,\mbox{MeV}$, 
allowing direct comparison to theoretical predictions 
as well as to other published measurements.

In order to reduce backgrounds in which one or more particles are missing 
or mismeasured in the detector, 
or which involve accidental activity, a momentum balance cut was imposed.
The square of the component of the total momentum of the decay perpendicular
to a line drawn from the target to the vertex was
required to be less than $300\,(\mevoc)^2$ for $\kleegg$ candidates.

\begin{figure}
\epsfig{file=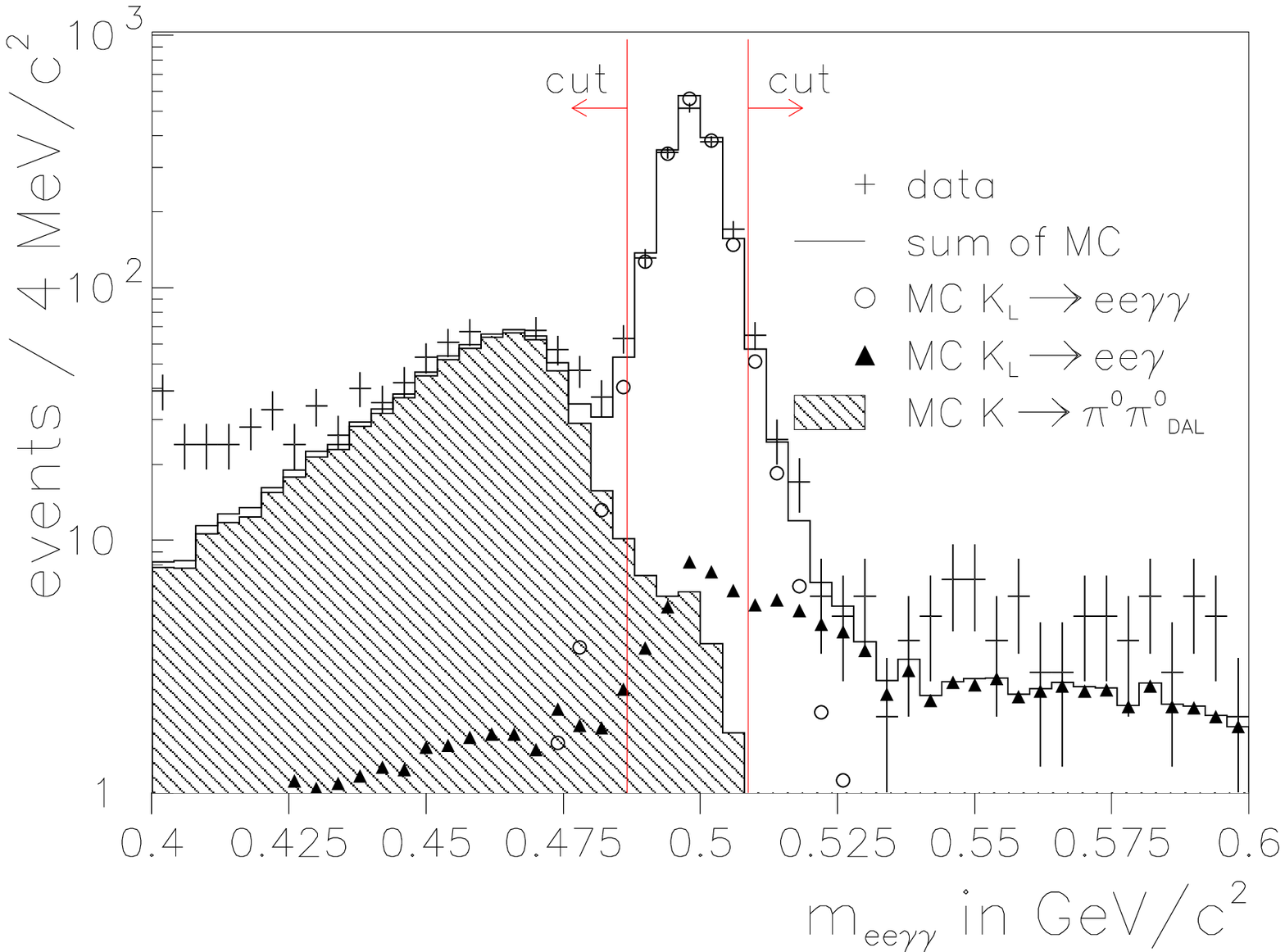,width=3.5in}
\caption{The $e^+e^-\gamma\gamma$ invariant mass distributions
for the data, the $\kleegg$ MC,
and the background MC samples. 
The MC distributions
use the normalization calculation
and predicted branching ratios (for $\kleegpg$)
or PDG branching ratio (for $\kpipid$).
}
\label{meegg}
\end{figure}

The invariant mass of the decay, $\meegg$, was calculated assuming
that the tracks were electrons.
This mass was required to be within $11\,\mevocc$ of 
the neutral-kaon mass.
Figure \ref{meegg} shows the distributions of $\meegg$,
after all other cuts,
for data, signal, and the two main backgrounds.
Values of $\meegg$ for $\kpipid$ decays tended to be less
than the kaon mass because one photon was missing.
The excess in data at very low $\meegg$ was from
$\kpipipid$ decays with three missing photons.
The $\kleeg$ decays often had a mass above the kaon
mass because most such decays 
had an extra, accidental photon.
The peak at the kaon mass in the $\kleeg$ distribution is
caused by events with external radiation.

\begin{figure}
\epsfig{file=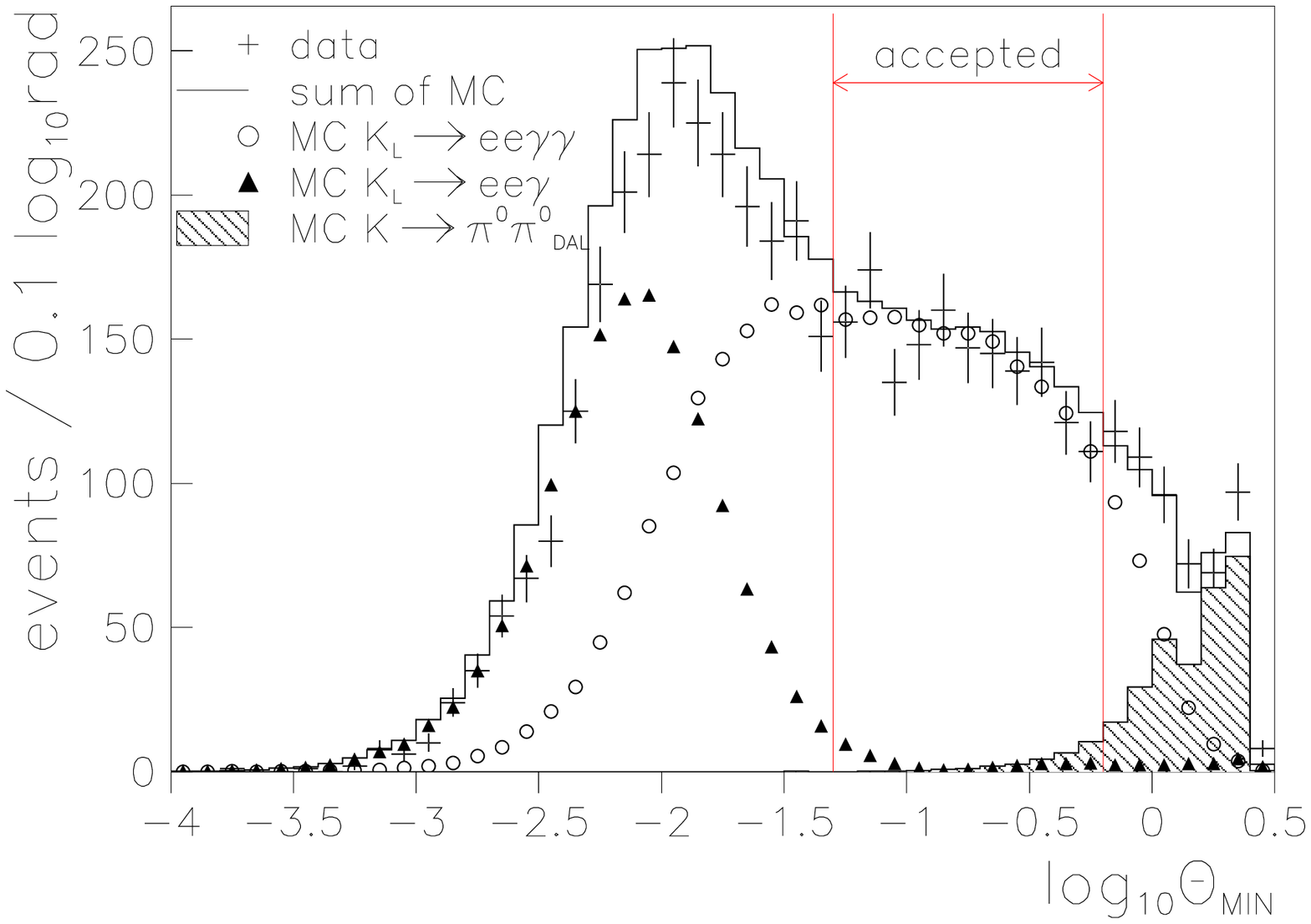,width=3.5in}
\caption{The $\log_{10}\mintheg$ distributions
for the data, the $\kleegg$ MC,
and the background MC samples.
The MC distributions
use the normalization calculation
and predicted branching ratios (for $\kleegpg$)
or PDG branching ratio (for $\kpipid$).
}
\label{mintheg}
\end{figure}

The distributions of minimum angle, 
$\mintheg$, between any photon and any electron
in the center of mass of the decay are plotted in Fig. \ref{mintheg}
after all other cuts.
The small angle events were dominated by $\kleeg$ with an externally
radiated photon.
The large angle events were dominated by $\kpipid$,
where the electron directions were relatively uncorrelated
with photon directions.
Signal events were required to have $0.05 < \mintheg < 0.63\,\mbox{rad}$.

\begin{figure}
\epsfig{file=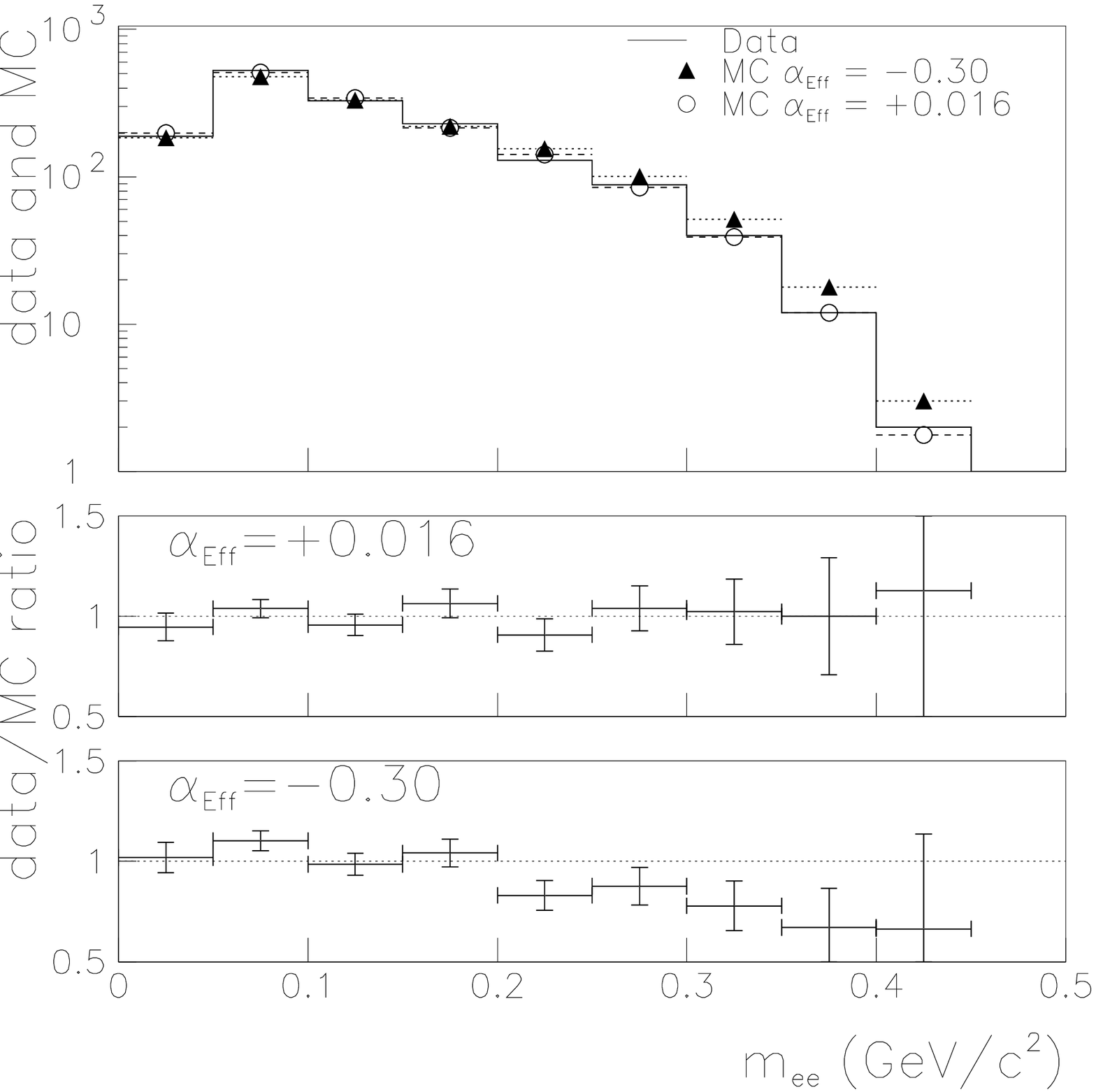,width=3.5in}
\caption{
The raw $e^+e^-$ invariant mass distribution for the data is shown followed
by data/MC ratios of $e^+e^-$ invariant mass($m_{ee}$) distributions,
for MC using $\aeff=0.016$ and $\aeff=-0.30$. The error bars are from 
data statistics. The MC distributions
were normalized to have the same integral as the data distribution.
}
\label{mee}
\end{figure}

There were 1,543 $\kleegg$ candidate events that satisfied all requirements.
The distribution of $e^+e^-$ mass in these events 
after background subtraction was used to determine the form factor.
However, since the MC did not include radiative corrections
to $\kleegg$, an effective parameter, $\aeff$, was
measured rather than the true $\akstar$.
We simulated the effect of different effective form factors in the MC
by reweighting $\kleegg$ and $\kleeg$ MC events with
a ratio of the BMS form factor squared over the generated form factor
(where $\aeff=-0.28$) squared.
The best fit to the data
was found at $\aeff=+0.016$ 
with $\chi^2=4.3$ for 9 degrees of freedom.
Fig. \ref{mee} shows the ratio of data to MC for this value
and for $\aeff=-0.30$.

The statistical uncertainty was found by using 22 subsets
of MC events, each
the same size as the data.  The spread of different $\aeff$ found by
using each subset in place of the data 
was taken as the statistical uncertainty, 0.083.
A sample of 12 million $\kpipipid$ decays was used to check
the MC simulation of the acceptance as a function of $m_{ee}$.  
A ratio of the data over
MC results in a slope of $0.038 \pm 0.096$ $\rm (GeV/c^{2})^{-1}$ and a 
$\pm 1 \sigma$ range of this
slope was fed back into the $\kleegg$ simulation, resulting in a systematic
uncertainty of $\pm0.032$ on $\aeff$.
Uncertainty in our ability to reconstruct events with $\kpipipid$ decays,
where the two (usually closely spaced) tracks both passed through one of
the beam regions of the first drift chamber led us to assign an
uncertainty to $\aeff$ of $\pm 0.027$.
Other systematic uncertainties, 
similar to those assessed below for the branching ratio, were evaluated, but
the results were negligible.
These systematic uncertainties are combined to give a total uncertainty of
$\pm0.042$ on $\aeff$.

The acceptance for $\kleegg$ with generated $\egkcm > 5\,\mbox{MeV}$
was found to be $0.984\%$ using
MC reweighted to have $\aeff=+0.016$.
A Monte Carlo simulation
that was normalized to the number of $K_L$ decays in the detector
predicted 31 background events
from $\kleeg$ (and $\kleegg$ with true $\egkcm < 5\,\mbox{MeV}$)
and 25 background events from $\kpipid$.
Thus, the branching ratio of $\kleeggc$ is 
$(5.84\pm0.15 {\rm ~(stat)})\times10^{-7}$.

Several sources of systematic uncertainty for the branching ratio measurement 
were considered.
The normalization calculation contributed uncertainties as mentioned 
previously.
A systematic uncertainty of $\pm0.49\%$ was assigned in order for the MC to 
match the relative excess of simulated events at small angles 
(Fig. \ref{mintheg}).
Another uncertainty was assigned
to account for uncertainty in the inefficiency
for tracks in the neutral-beam region of the drift chambers (the
beam region was $0.965\pm0.028$ efficient),
by reweighting MC events (signal, backgrounds, and normalization) 
based on the number of tracks in the beam region in each event;
the resulting $0.49\%$ drop in branching ratio was
taken as a symmetric uncertainty.
Varying $\aeff$ by our combined uncertainty of $\pm0.093$
changed the acceptance and $\kleeg$ background,
and the resulting $\pm2.25\%$ change in branching ratio was taken 
as an uncertainty.
Finally, uncertainties due to inaccuracies in the MC simulation
of the detector were estimated by varying one cut at a time over a 
reasonable range
simultaneously for data, signal, backgrounds, and normalization.
The most significant of these uncertainties were from the
transverse-momentum cut, the cut on the opening angles
between the the two tracks in the lab frame, and the $\meegg$ cut.
When added in quadrature, all of the cut variations contributed
a systematic uncertainty of $\pm3.92\%$.
The combined systematic uncertainty on ${\mathcal B}(\kleeggc)$ was 
$\pm5.44\%$.

MC numerical integration of the tree-level QED partial width,
using the BMS form factor with $\akstar=+0.016$ and 
${\mathcal B}(K_L\to\gamma\gamma)$ = $(5.86\pm0.15)\times10^{-4}$ \cite{PDGN},
predicts ${\mathcal B}(\kleeggc)$ = $(5.70\pm0.14)\times10^{-7}$,
based on ${\mathcal B}(\kleeg)$ = $(9.1\pm0.5)\times10^{-6}$,
in agreement with the measurement.

In summary,
we have determined the branching ratio of $\kleegg$,  with an infrared 
threshold of $\egkcm > 5\,\mbox{MeV}$, to be
$(5.84 \pm 0.15 {\rm ~(stat)} \pm 0.32 {\rm ~(sys)})\times 10^{-7}$.
This calculation uses an effective form factor, which does not include
effects from radiative corrections, with
$\aeff = +0.016\pm0.083{\rm ~(stat)} \pm 0.042{\rm ~(sys)}$.
This measurement is an improvement in precision over previous measurements,
and agrees with the predicted value at tree level.


%
%
We gratefully acknowledge the support and effort of the Fermilab
staff and the technical staffs of the participating institutions for
their vital contributions.  This work was supported in part by the U.S. 
Department of Energy, The National Science Foundation and The Ministry of
Education and Science of Japan.

\end{document}